\documentclass[12pt]{ar}
\usepackage[english]{babel}
\usepackage{psfig,graphicx}
\usepackage{supertab}

\onecolumn
\begin{document}

\title{Evolutionary stage of the spectral variable BD\,+48${\rm ^o}$1220\,=\,IRAS\,05040+4820}

\author{V.G.~Klochkova, E.L.~Chentsov, N.S.~Tavolganskaya \& V.E.~Panchuk}

\date{\today}	     

\institute{Special Astrophysical Observatory RAS, Nizhnij Arkhyz,  369167 Russia}

\abstract{Based on high-resolution observations (R\,=\,60000 and 75000),
we have studied the optical spectral variability of the star
BD+48${\rm ^o}$\,1220, associated with the IR source IRAS\,05040\,+\,4820. We
have measured the equivalent widths of numerous absorption lines of
neutral atoms and ions within the region 4500--6760\,\AA, as well
as the corresponding radial velocities. We use model atmospheres method
to determine the effective temperature T$_{eff}$\,=\,7900\,K, surface gravity
$\log g$\,=\,0.0, microturbulence velocity $\xi_t$\,=\,6.0,
and the abundances for 16 elements. The metallicity of BD+48${\rm ^o}$\,1220
differs little from the solar one: [Fe/H]=$-0.10$\,dex. The main
peculiarities of the chemical content of the star are a large helium
excess, derived from the HeI\,5876\,\AA{} absorption, [He/H]\,=\,+1.04,
and the oxygen excess, [O/Fe]\,=\,+0.72\,dex. The carbon overabundance is
small, [C/Fe]\,=\,+0.09 dex, and the ratio [C/O]$< 1$. We obtained an
altered relation for the light-metal abundances: [Na/Fe]\,=\,+0.87\,dex
with [Mg/Fe]\,=\,$-0.31$\,dex. The barium abundance is lowered,
[Ba/Fe]\,=\,$-0.84$\,dex. We concluded that the selective separation of
elements onto dust grains of the envelope is probably efficient. The
radial velocity of the star measured from photospheric absorption lines
over three years of observations varies in the interval $V_r = -(7\div 15)$\,km/s.
Time variable differential line shifts have been revealed. The
entire set of available data (the luminosity $Mv\approx -5^m$, velocity
$V_{lsr}=-20$\,km/s, metallicity [Fe/H]\,=\,$-0.10$, and peculiarities of
the optical spectrum and chemical composition) confirms the status of
BD+48${\rm ^o}$\,1220  as a massive post-AGB star with He-- and O-- excesses
belonging to the disk population.}

\authorrunning{Klochkova et al.}
\titlerunning{Spectroscopy of the spectral variable BD\,+48${\rm ^o}$1220}

\maketitle

\section{Introduction}

The 6\,m telescope of the Special Astrophysical Observatory (SAO) has been
used for spectroscopy of highly luminous stars with large
IR--excesses on the asymptotic giant branch (AGB) or in the post-AGB
stage. These objects are a transition stage to becoming a planetary nebula
(for this reason, they are called protoplanetary nebulae, or PPN). PPN
have passed through a long evolution: hydrogen and helium burning both in
the core and in layer sources, mixing and mass loss via stellar wind at
the AGB stage. The current theory of the evolution of intermediate-mass
stars (see, for example, [\cite{Herwig}]) gives grounds to expect altered
elemental abundances in the surface layers of these evolved objects. Our
program is mainly aimed at determining the fundamental parameters of stars
that likely belong to this class, identifying their evolutionary stage,
and searching for evolutionary variations of their chemical compositions.
In addition, we are searching for spectral variability, and also carry out
detailed study of the velocity fields in their atmospheres and envelopes,
for comparison with those in supergiants and hypergiants and to reveal
possible binarity in these objects. The basic results obtained in the
program are summarized in [\cite{rev, KPS}]. Papers [\cite{Egg, V510}],
which initiate our study of the spectroscopy of bipolar nebulae and of the
peculiar hypergiant IRC\,+10420, whose evolution status has become fairly
firm only recently (in particular, owing to our results [\cite{IRC1,
IRC2}]), are also of interest in this context. Here, we present the
results for our spectral observations of the thus far poorly studied star
BD\,+48${\rm ^o}$1220 (SAO 40039, LSV\,+48${\rm ^o}$\,26)--the optical
counterpart of the IR-source IRAS\,05040+4820 (from now on -- IRAS\,05040).
BD\,+48${\rm ^o}$1220 displays a double--peaked spectral energy
distribution (SED): the IRAS catalog [\cite{IRAS}] suggests somewhat high
IR fluxes $f_{12}$=0.25, $f_{25}$=7.20, $f_{60}$=20.20, $f_{100}$=11.00.

The evolutionary stage of BD\,+48${\rm ^o}$1220 remains unclear. The star
is situated fairly close to the plane of the Galaxy
(b\,=\,4$\lefteqn{.}^m8$, l\,=\,159$\lefteqn{.}^m8$), which may indicate
that it is young and probably belongs to the disk population. The high
luminosity of BD\,+48${\rm ^o}$1220 and its two--peaked SED, which
indicates the presence of circumstellar matter ejected in the course of
its previous evolution, make it a good candidate post--AGB star, and Fujii
et al. [\cite{Fujii}] recently did, in fact, assign BD\,+48${\rm ^o}$1220
to this stage. Based on multi--color optical photometric observations for
a sample of high-luminosity stars, Fujii et al. [9] obtained for
BD\,+48${\rm ^o}$1220 B\,=\,10$\lefteqn{.}^m1$ and
V\,=\,9$\lefteqn{.}^m65$, and determined its spectral type to be
Sp\,=\,A4Ia.

Using the 6\,m telescope in combination with an echelle spectrograph
during several observing seasons, we have obtained the first high
resolution optical spectra of BD\,+48${\rm ^o}$1220, and detected substantial
variability of the H\,I and metallic line profiles [\cite{IBVS}]. Here, we
present the results of a detailed spectral study of  BD\,+48${\rm ^o}$1220
using model atmospheres and synthetic spectra methods, as well as a
detailed analysis of the pattern of radial velocities in the atmosphere of
the star and the circumstellar space. Section\,\ref{methods} briefly
describes our methods and the reduction of the spectral data, the
peculiarities of the spectrum of BD\,+48${\rm ^o}$1220, and our results
for its fundamental parameters. In Section\,\ref{discuss}, we discuss our
results and the evolutionary status of the star.

\section{Observations, data reduction and analysis}\label{methods}

\subsection{Echelle spectroscopy with the 6\,m telescope}

The spectroscopy of BD\,+48${\rm ^o}$1220 was carried out with the 6\,m
telescope of the SAO. All the spectra were obtained at the Nasmyth focus
with the NES echelle spectrograph [\cite{nes}] equipped with a
2048\,$\times$\,2048 CCD. In combination with the image slicer
[\cite{slicer}], the NES spectrograph provides a spectral resolution of
R\,=\,60000. Table\,1 presents the dates of the observations, the
integration times, the recorded spectral interval and S/N ratio. The data
were extracted from the two-dimension echelle spectra using the modified
[\cite{reduc}] ECHELLE procedure of the MIDAS package running under an OS
Linux. Hints of cosmic rays were removed via median averaging of pairs of
consecutive spectra. The wavelength scale was calibrated using spectra
produced by a Th-Ar lamp. The subsequent reduction, including photometric
and position measurements, was carried out with the DECH20 code
[\cite{gala}]. For each spectrogram, the position zero point was
determined via a standard calibration using the positions of ionospheric
night-sky emission lines and telluric absorption lines observed against
the background of the object's spectrum. The accuracy of the velocity
measurements in the spectra obtained with the NES spectrograph is better
than 1\,km/s for a single line.

\subsection{Main peculiarities of the optical spectrum}

{\it Emission features.} As was already noted [\cite{IBVS}], the optical
spectrum of BD\,+48${\rm ^o}$1220 displays numerous emission components.
The H$\alpha$ line demonstrates a complex two--component emission profile,
which is time-variable, as can be seen from Fig.\,1. Here, the horizontal
and vertical axes plot radial velocity $V_{\odot}$ and intensity $I$ (note
that the continuum level is specified to be 100). The core of the H$\beta$
line profile is also distorted by variable emission features. Variable
emission components are also observed in the SiII, FeI, and FeII metal
lines. To illustrate this fact, Fig.\,2 presents the behavior of the
FeII\,(55) 5534\,\AA{} line for several dates. In the spectrum obtained on
March 8, 2004, the emission at $V_r \approx -20$\,km/s is also visible in
the profiles of the FeII 40, 46 and 74 multiple lines.

For a star whose effective temperature is below 8000\,K, the
HeI\,5876\,\AA{} neutral helium absorption line is unexpectedly strong;
its equivalent width, $W_{\lambda}(5875)$=75\,m\AA{}, is measured reliably
in all available spectra. The absorption nature of the line indicates its
photospheric origin, as is also confirmed by the fact that the radial
velocity corresponding to the position of the line is consistent with the
velocity obtained from other weak absorption features (see Table\,1 in the
Appendix).

{\it Interstellar features and color excess.} A detailed examination
of the high-resolution spectra shows that the NaI  double resonance
D--lines in the spectrum of BD\,+48${\rm ^o}$ 1220 display complex
profiles with several absorption components. Figure\,3 presents the
D1--line profiles for various observing times. Here, we can distinguish
components whose positions correspond to the following velocities: three
components in the velocity interval $V_{\odot}=-20$ to $-30$\,km/s and a
strong and broad component around $V_{\odot}\approx -2$\,km/s
($V_{lsr}$=$-8$\,km/s). According to M\"unch [\cite{Munch}], interstellar
D2 NaI lines are observed in a direction close to the position of
BD\,+48${\rm ^o}$1220, whose position corresponds to sets of velocities
$V_{lsr}\approx$$-4$ to $-11$ and $-24$ to $-33$\,km/s. In the CO--band
survey of the Milky Way [\cite{Dame}], at a longitude of
l\,=\,160$\rm ^o$, two intervals of the interstellar gas velocities are in
a good consistency with the data of Munch [15]: $V_{lsr}$=$-10$ to +7, and
$-17$ to $-30$\,km/s.

The spectrum of BD\,+48${\rm ^o}$1220 is particularly remarkable for its
numerous absorption features identified with diffuse interstellar bands
(DIBs). Table\,2 presents a list of such features and their equivalent
widths $W_{\lambda}$, and the heliocentric velocities corresponding to the
positions of these features are given in Tabl.\,1 in the Appendix. The
average velocity $V_{\odot}$ derived from the interstellar bands presented
in Tabl.\,3 is consistent with that determined from the main component of
the D--lines, $V_{\odot}\approx -2$\,km/s ($V_{lsr}=-8$\,km/s), which
confirms its interstellar origin (Fig.\,4).

Previously [\cite{IBVS}] the absolute luminosity $M_v\approx -5^m$
(luminosity type Ib) was obtained for BD\,+48${\rm ^o}$1220, which
corresponds to a distance of d$\approx$5\,kpc, taking into account
interstellar absorption. Using the calibration of Herbig [\cite{Herbig}]
for $W_{\lambda}(5780)$--$E(B-V)$, we obtain from the measured value
$W_{\lambda}$=227\,m\AA{} for the 5780\,\AA{} band the color excess
$E(B-V)\approx 0.4^m$, which leads us to the interstellar absorption $A_v
\approx 1.2$. This suggests that the color excess of BD\,+48${\rm ^o}$1220
is due to interstellar absorption. However, according to Neckel and Klare
[\cite{Neckel}], the absorption in the disk of the Galaxy in the direction
towards the object in longitude l\,=\,160$^o$ already increases to $A_v
\ge 1^m$ at a distance of 1\,kpc. Thus, the formal application of Herbig's
calibration substantially decreases the distance to the object. In the
case of a star with an envelope, some of the color excess can be due to
absorption in the circumstellar region; therefore, the distance estimated
from the amount of absorption can prove to be incorrect.

In the context of the star's color excess, it is of interest to consider
the optical polarizations of 26 PPN candidates determined by Parthasarathy
et al. [\cite{Part}], including BD\,+48${\rm ^o}$1220. According to these
data, BD\,+48${\rm ^o}$1220 belongs to a very small subgroup of stars with
high polarizations: within the entire $BVRI$ range, the degree of
polarization is $p>$\,3\%. The available wideband polarimetric data do not
provide a firm description of the behavior of the polarization with
increasing wavelength, hindering identification of the polarization
mechanism. However, the high degree of polarization, which substantially
exceeds the interstellar value for a color excess $E(B - V) \approx 0.4$,
suggests that the radiation is polarized in the circumstellar environment
of the star. In addition, measurements made on different dates display
variability of the polarization: for example, in $V$, it varies from
$p$=3.15\% to 3.61\%. This variability also confirms the "stellar" origin
of the polarization, which, as in some other PPNs (IRAS\,04296+3429,
IRAS\,08005$-$2356 [\cite{Tram}]), could be associated with the
asphericity and dynamics of the circumstellar envelope.

\subsection{Parameters of the stellar atmosphere}

To determine the main model parameters of the atmosphere, such as the
effective temperature and gravity, and also to calculate the chemical
composition and synthetic spectra, we used the grid of model stellar
atmospheres [\cite{Tsymbal}] calculated in a hydrostatic approximation
assuming local thermodynamic equilibrium (LTE) for various metallicities.
The most difficult problem in calculating the chemical composition of a
star is always fixing its basic parameters -- the effective temperature
$T_{eff}$ and gravity $\log g$. For objects with uncertain evolutionary
statuses, and hence uncertain reddening, it is difficult to apply
photometric data to determine the effective temperature. Profiles of the
hydrogen Balmer lines are widely used to determine the parameters of
A--stars. However, the profiles of these lines in the spectrum of
BD\,+48${\rm ^o}$1220 are distorted by variable emission. Therefore, we
determined $T_{eff}$ of the star using our spectral data, from the
condition that the abundance from neutral iron be independent of the
excitation potential of the FeI lines used. The gravity was determined
from the condition that the iron atoms be in ionization balance, and the
microturbulence velocity $\xi_t$ from the condition that the iron
abundance be independent of the line intensities. An additional criterion
for the reliability of the method is the absence of this dependence for
other elements represented in the spectra by numerous lines (CrII, TiII).
In addition, if $\xi_t$ is determined confidently, no dependence of the
individual abundances on the equivalent widths of the lines used for the
calculations should be seen.

The typical accuracies of the model parameters (on average) for a star
with an effective temperature of about 8000 are $\Delta
T_{eff}\approx$200\,K, $\Delta \log g \approx$ 0.5\,dex, $\Delta \xi_t
\approx$\,0.5\,km/s. The high luminosity of BD\,+48${\rm ^o}$1220 is also
confirmed by the very large equivalent widths of the ionized silicon SiII
6347, 6371\AA{} lines: $W_{\lambda}$=516 and 490\,mA, respectively
(Fig.\,5). According to Venn [\cite{Venn95}], the equivalent widths of the
SiII\,6347\,\AA{} lines in the spectra of stars in a sample of A0--F0
supergiants are a factor of 1.5--2 lower than in the spectrum of
BD\,+48${\rm ^o}$1220. The application of a static plane-parallel model
assuming LTE--approach to the unstable atmosphere of an A--supergiant may
seem doubtful. This problem is considered in detail by Venn in [\cite{Venn95}];
however, Venn  concluded here that blanketed Kurucz LTE models
provide the best available fits for high luminous A--stars. Checking for a
good correlation between the observed and synthetic spectra provides a
self-consistency test for obtained parameters. To check the reliability of
the model atmosphere parameters, we compared the observed spectrum with a
synthetic spectrum with the solar chemical composition and the model
parameters $T_{eff}$=7900\,K, $\log$=0.5, and $\xi_t$=6.0\,km/s calculated
using the code [\cite{Tsymbal}]. This comparison (Fig.\,5) shows reasonable
consistency, with the exception of the strong SiII lines. Table\,4 presents the
results of calculations of the chemical composition with the model
$T_{eff}$=7900\,K, $\log g$=0.0, $\xi_t$=6.0\,km/s. The elemental
abundances (X) calculated for individual lines with the parameters
$T_{eff}$=7900\,K, $\log g$=0.5, $\xi_t$=6.0\,km/s km/s are given in
Table\,2 in the Appendix, together with the oscillator strengths $\log gf$
and other atomic constants from the VALD database [\cite{VALD1, VALD2}].
Note that this table contains only the selected spectral lines used to
derive the model parameters and elemental abundances.

\section{Discussion of the results}\label{discuss}

\subsection{Kinematics of the stellar atmosphere}

All our measurements of the radial velocity $V_{\odot}$ for various
observing moments are presented in Tabl.\,3, which shows that all the
spectral features display time variability in $V_{\odot}$. For the period
of our observations, the minimum variations in $V_{\odot}$ were obtained
for weak ($r \rightarrow 1$) absorption lines of metals: $V_{\odot}$=$-7$
to $-15$\,km/s. The velocities derived from individual lines are given in
Table\,1 in the Appendix. Figure\,4 presents the distribution of the
radial velocities $V_{\odot}$ measured from lines with different depths
for one of the spectra (January 10, 2004). Here, the vertical and
horizontal axes plot radial velocity and line depth (for the strongest
lines, the depths on this scale are close to 100).

The systemic velocity $V_{sys}$ is a fundamental parameter in the
kinematic pattern shown by the lines of a star. But 
observations of BD\,+48${\rm ^o}$1220 at radio wavelengths [\cite{Wout1,
Wout2}] have not revealed any spectral features that can be used to fix
the value of $V_{sys}$. Therefore, as a first approximation, we adopted
the average of the velocities for most weak absorption features:
$V_{sys}\approx -13$\,km/s ($V_{lsr}\approx -20$\,km/s). This value is
consistent with the velocities of H\,II regions in the local volume of the
Galaxy in the direction towards BD\,+48${\rm ^o}$1220:
$V_{lsr}$(HII)$\approx -20$\,km/s [\cite{Georg}]. Brand and Blitz
[\cite{Brand}] found the velocity $V_{lsr}$(HII)$\approx -20.5$\,km/s in
the same direction, for a distant region of the Galaxy at a distance of
$d$=5.2\,kpc. Recall that, according to our estimate (see Section\,2.2),
the distance to the star is $d\approx$5\,kpc. The absorption core of the
H$\beta$ in the spectrum of BD\,+48${\rm ^o}$1220 is free of emission, and
the velocity derived from the H$\beta$ core for two dates when it is
detected is consistent with the value derived from the weak absorptions.
The cores of the absorption lines of ions (FeII, CrII, etc.) indicate
larger variations. The maximum displacement relative to weak lines was
detected for the core of the H$\alpha$ line: at two epochs, this line is
displaced short-ward of the systemic velocity by more than 10\,km/s
(Fig.\,4). As follows from Fig.\,4, the velocity corresponding to the
position of the H$\alpha$ absorption is consistent with the position
derived from the short-wavelength component of the NaI D--line,
$V_{\odot}\approx -28$\,km/s. Thus, in addition to the interstellar
constituent, the short--wavelength component of D--lines NaI may also have
a circumstellar (wind) contribution. Regrettably, the obtained spectra are
insufficient for us to draw more definite conclusions. Obviously, further
spectral monitoring and spectropolarimetry of the star are needed.

\subsection{Chemical abundances pattern}

Table\,4 presents the adopted model atmosphere parameters $T_{eff}$, $\log
g$, $\xi_t$ and the average elemental abundances relative to iron, [X/Fe].
The chemical composition of the solar photosphere, relative to which we
will consider the abundances in BD\,+48${\rm ^o}$1220, is taken from
[\cite{Lodders}]. All the calculations of the chemical composition were
made using codes developed by Shulyak et al. [\cite{Tsymbal}] and adapted
for a PC running with OS Linux. The plane parallel LTE--models were
calculated using the codes described in [\cite{Tsymbal}]. Corrections
for hyperfine structure and isotopic shifts for the NiI, MnI, and BaII lines were not
taken into account. The scatter of the abundances obtained from the set of
lines is small: the $rms$ deviation essentially does not exceed 0.3\,dex
(see Table\,4). When determining the model parameters, we used lines with small
and medium intensities with $W_{\lambda}\le$0.25\,\AA{}, since the
approximation of a stationary plane-parallel atmosphere could be
inadequate for the strongest spectral features. In addition, some strong
absorption lines could be distorted by the effect of the circumstellar
envelope; with insufficient spectral resolution, the intensity of envelope
components is added to that of lines formed in the atmosphere.

The abundances were calculated for an extended set of lines: for some
elements (MgI, TiII, CrII, FeII), we included lines with equivalent widths
exceeding the above limit. As follows from Table\,2 in the Appendix 
the abundances determined from strong lines do not differ systematically
from those derived from lines with low and medium intensities. This is
illustrated well by the results for silicon: the abundances derived from
the weak SiII\,4621\,\AA{} line and from two very strong SiII\,6347 and
6371\,\AA{} lines coincide. Let us consider the detailed abundances,
combining them into groups according to the type of elemental synthesis.

{\it Light elements.} As should be expected for a high-luminosity star, the
abundances of some light elements are altered from their initial values.
First and foremost, the helium excess derived from the HeI\,5876\,\AA{}
line stands out. Figure\,6 compares the observed spectrum (for March 8,
2004) and the synthetic spectrum calculated for the model
$T_{eff}$=7900\,K, $\log g$=0.5, $\xi_t$=6.0\,km/s and the chemical
composition from Table\,4. The two spectra are in reasonable agreement.

Since the observed LiI\,6707\,\AA{} line is very weak, the derived lithium
excess is subject to considerable uncertainty.

We find an appreciable oxygen excess [O/Fe]=+0.72, with an almost normal
carbon abundance [C/Fe]=+0.09; the ratio $C/O <$1. Figure\,7 compares the
observed spectrum (January 10, 2004) with the synthetic spectrum. The
essential oxygen excess indicates that BD\,+48${\rm ^o}$1220 cannot be a
massive supergiant. Analyses of spectra of massive stars [\cite{Venn93}]
shows that, in accordance with theoretical predictions, the evolution of a
massive star leads to a deficit of oxygen, which is reprocessed into
nitrogen in the course of the CNO cycle. Unfortunately, the detected
spectral interval was limited to $\lambda$=6760\,\AA{}; therefore, no
nitrogen lines were measured, although the N abundance is essential for
determining the evolution stage of the star.

As we can see from Table\,2 in the Appendix, the sodium abundance was
determined from the weak NaI 5682, 5688, and 6160\,\AA{} lines, for which
corrections due to deviations from LTE [\cite{Takeda, Zhao}] are the
smallest. Therefore, the derived sodium excess, denotes the dispersion of
the abundance obtained for a given number of lines n. The elemental
abundances for the solar photosphere are taken from [\cite{Lodders}].
Sodium excess, [NaI/Fe]=+0.87, could basically be the result of sodium
synthesis in the NeNa cycle, which occurs simultaneously with hydrogen
burning in the CNO cycle [\cite{Denis1, Denis2}].

{\it Iron-peak elements.} The iron abundance, generally taken to be the
metallicity, in the atmosphere of BD\,+48${\rm ^o}$1220 differs little
from the solar value: $\log \varepsilon(\rm FeI,FeII)$=7.37. Sufficiently
trustworthy abundances estimates for vanadium and chromium, which belong
to the iron group, also display only small deviations from the normal
values: [VII,CrI,CrII/Fe]=+0.10. A metallicity close to the solar value is
consistent with the system's heliocentric radial velocity
$V_{\odot}\approx -13$\,km/s, adopted in [\cite{IBVS}], which is typical
for stars in the Galactic disk. However, the manganese, and especially
nickel, abundances deviate substantially from the normal values. While the
anomalous manganese abundance could be explained by our lack of inclusion
of hyperfine structure of the lines and relatively low accuracy due to the
small number of lines, the nickel overabundance [NiI/Fe]=+0.74 has been
firmly determined from three medium intensity lines. Therefore, the nickel
excess appears to be real.

{\it Heavy metals.} The large barium deficit relative to iron,
[Ba/Fe]=$-0.84$, is typical of the atmospheres of supergiants. A
deficiency of $s$-process elements in the atmospheres post-AGB stars is
observed much more often than an excess [\cite{rev, Winck}]. The observed
lack of observed manifestations of the dredge-up of heavy metals is
probably real, rather than resulting from systematic errors in the
analysis of spectra of supergiants using model atmospheres. The presence
or absence of $s$--process elements overabundance is probably related to
the initial mass of the star and the mass-loss rate in the AGB stage,
which specify the evolution of a given star and the mass of its core.
Modelling of the process of third dredge-up [\cite{Herwig2}] indicates
that the efficiency of the dredge-up increases with the growth of the core
mass (and hence of the initial mass) of a post--AGB star. Calculations
made by Herwig [\cite{Herwig3}] also show that the efficiency of the
dredge-up increases if penetrating convection (overshooting) at the base
of the convective zone is taken into account.

\subsection{The separation of elements in the envelope}

It is known that the selective separation of elements can provide an
efficient mechanism for creating anomalous abundances of elements in
atmospheres of stars with gas--dust envelopes. The star studied can
undergo a stage of intense mass exchange between the atmosphere and the
circumstellar gas--dust envelope; generally, the  pattern of abundances is
consistent with the dependence on the condensation temperature. Another
argument in favor of separation is the enhanced zinc abundance:
[Zn/Fe]=+0.44.
The zinc abundance is a good test, since it is only slightly
susceptible to selective condensation and does not vary in the course of
the nuclear evolution of the given star. As was shown by Sneden et al.
[\cite{Sneden}], within a large interval of metallicity, the zinc
abundance corresponds to the behavior of iron, [Zn/Met]=+0.04. This
conclusion was clarified by the recent results of Mishenina et al. [\cite{Mish}],
which showed that zinc followed the metallicity within the total studied
interval [Fe/H]=$-0.5$ to $-3.04$. For stars of the disk population, a slight
increase of the zinc abundance with decreasing metallicity from [Fe/H]=0  
to [Fe/H]=$-0.6$\,dex has been obtained [\cite{Chen}].

Additionally, we expect that elements with similar condensation
temperatures $T_{cond}$ should behave in the same way. For example,
according to [\cite{Lodders}], $T_{cond}$ for Ca and Sc are close. Indeed, as we
can see from Table\,4, these elements have very similar relative abundances
in BD\,+48${\rm ^o}$1220. We conclude that the anomalous chemical
composition can be explained by the selective condensation of atoms on
dust grains in the circumstellar environs of the star. However, the fact
that the iron abundance is close to the solar value indicates that
anomalies of the elemental abundances introduced by condensation on dust
grains are not very substantial, since iron is efficiently depleted onto
dust grains. In addition, the behavior of some elements is inconsistent
with the dependence on the condensation temperature: for example, Ni,
whose $T_{cond}$ is close to that of Fe and V, displays an appreciable
excess abundance. The situation is even more complicated by the fact that,
apart from being affected by selective condensation, the abundances of
some elements (C, O, Ba), can also vary due to nuclear
processes in the course of the evolution of the star.

Chemical compositions similar to that obtained for BD\,+48${\rm ^o}$1220
are not unique for post--AGB stars. The elemental abundances in the
atmospheres of PPNs can vary greatly (see, for example, the summaries in
[\cite{rev, Winck}]). Previously studied PPN candidates include
stars with similarly anomalous chemical compositions. For example,
HD\,331319 (the optical counterpart of IRAS 19475+3119) also displays a
metallicity close to the solar value, an excess of He, O, and light metals
(Na, Si), and a deficit of Ti and Ba [42]. As in the case of BD\,+48${\rm
^o}$1220, the neutral He\,5876\,\AA{}) absorption is also observed
($W_{\lambda}$=38\,m\AA{} in the spectrum of the F--supergiant HD\,331319.
Klochkova et al. [\cite{331319}] concluded that the He--lines were
photospheric origin, and hence the He--excess due to the dregde-up of
nuclear products synthesized during the evolution of the star was real.
HD\,161796 [\cite{331319, LBL}] (the optical counterpart of
IRAS\,17436+5003) is another similar star. The A--supergiant HD\,133656
(IRAS\,15039$-$4806) [\cite{Oudm}] also displays a similar chemical abundances
pattern; however, due to its decreased metallicity ([Fe/H]$\le -0.7$), this
star is assigned to an older population of the Galaxy.

\subsection{Evolutionary status}

Determining the evolutionary status of A--supergiants in the Galactic field
is a nontrivial problem, since stars with substantially different ages and
masses observed at very different evolutionary stages can display the same
fundamental parameters, $T_{eff}$ and $\log g$. The same region in the
HR--diagram is occupied by post--AGB stars that are evolving from the AGB
stage to planetary nebulae and young, massive supergiants evolving from
the main sequence to the red supergiant stage. This problem of ``spectral
mimicry'' is especially challenging, for example, for objects such as the
peculiar hypergiant IRC+10420 [\cite{IRC1, IRC2}]. For BD\,+48${\rm
^o}$1220, the problem becomes even more complicated due to the fact that
this supergiant is situated close to the Galactic plane, and the radial
velocity pattern characteristic of supergiants is superimposed with
differential shifts. According to the empirical chronological sequence
[\cite{Lewis}], the absence of the standard features inherent to PPNs
(maser lines of OH, H$_2$O, CO, etc.) in the radio spectrum of
BD\,+48${\rm ^o}$1220 provides evidence that the star is close to the
planetary nebula phase, as is confirmed by its fairly high effective
temperature.

It is obvious that, when considering the evolutionary status of a star,
the peculiarities of its chemical composition are critical. In the
HR--diagram [\cite{Blocker}], the absolute magnitude $M_v \approx -5^m$ and
$T_{eff}$=7900\,K obtained for BD\,+48${\rm ^o}$1220 imply a fairly high
mass for the star: $\mathcal{M} \approx 4 \mathcal{M}_{\odot}$.
This value suggests that hot bottom burning (HBB) may be efficient
[\cite{Herwig}], as is common for more massive AGB stars (the mass limit
$\mathcal{M} \ge 4 \mathcal{M}_{\odot}$ depends on the metallicity). The
possibility of efficient HBB is also supported by the peculiarities of the
chemical composition of BD\,+48${\rm ^o}$1220, first and foremost, the
lithium excess [Li/Fe]=+0.62. Unfortunately, due to the weakness of the
LiI line, this excess has been determined with large uncertainty. However,
the O--excess, the low C/O value, and the Na--excess in the
atmosphere of BD\,+48${\rm ^o}$1220 are determined confidently, and
support the efficiency of the HBB mechanism. The nitrogen abundance could
be critical for verifying this hypothesis, making spectroscopy at longer
wavelengths important.

\section{Conclusions}

Based on our observations with high spectral resolution (R=60000 and
75000), we have studied the optical spectrum of the A--supergiant
BD\,+48${\rm ^o}$1220, the optical counterpart of the IR--source
IRAS\,05040+4820, in detail. At wavelengths from 4500 to 6760\,\AA{},
numerous absorption lines of neutral atoms and ions have been identified,
and their equivalent widths and corresponding radial velocities have been
measured. Using model atmospheres, we have obtained the effective
temperature ($T_{eff}$=7900\,K), gravity ($\log g$=0.0), microturbulence
velocity ($\xi_t$=6.0), and the abundances of 16 chemical elements. The
star displays close to solar metallicity: [Fe/H]=$-0.10$\,dex. The
abundances of the iron--group elements V and Cr display equally small
deviations from their normal values: [VII,CrI,ICrI/Fe]=+0.10. The main
feature of the chemical composition of the star is the large
He-- excess derived from the HeI\,5876\AA{} line, [He/H]=+1.04,
and the substantial O--excess [O/Fe]=+0.72\,dex.
The carbon excess is small: [C/Fe]=+0.10\,dex.
The relations between the abundances of light metals are anomalous:
[Na/Fe]=+0.87\,dex for [Mg/Fe]=$-0.31$\,dex, while [Na/Mg]=\,+1.18\,dex.
The Ba abundance  is appreciably low: [Ba/Fe]=$-0.84$\,dex.
The luminosity of the star and the peculiarities of its
chemical composition (the excess of He, Li, and O with [C/O]$<$1)
suggest that BD\,+48${\rm ^o}$1220 is related to more massive
post--AGB stars, in which nuclear reactions occur at the base of the hot
convective envelope in the AGB stage (hot bottom burning). In addition,
some anomalies in the elemental abundances could be due to selective
condensation onto dust grains of the envelope. This hypothesis is
supported by the enhanced Zn abundance [Zn/Fe]=+0.44 and the similar
deficits of  Ca and Sc.

During the three years of our observations, the radial velocity of the
star derived from photospheric absorptions  varied in the interval
$V_{\odot}$=$-7$ to $-15$\,km/s. The available data
($M_v \approx -5^m$, $V_{lsr}\approx -20$\,km/s,
[Fe/H]=$-0.10$, the peculiarities of the optical spectrum and chemical composition)
confirm the status of BD\,+48${\rm ^o}$1220 as a helium and oxygen-rich
post--AGB star situated in the disk of the Galaxy.

\section*{Acknowledgements}

This work was supported by the Russian Foundation for Basic Research
(project code 05--07--90087), the program of basic research held by the
Presidium of RAS ``The origin and evolution of stars and the Galaxy'',
program held by the Section for Physical sciences of RAS ``Extended objects
in the Universe''. This publication is based on work supported by Award
No.RUP1--2687--NA--05 of the U.S. Civilian Research \& Development
Foundation (CRDF). In our study we used SIMBAD, ADS and VALD databases.

\newpage

\begin{figure}[t]	      		      
\includegraphics[angle=0,width=0.6\textwidth,bb=160 30 550 790,clip]{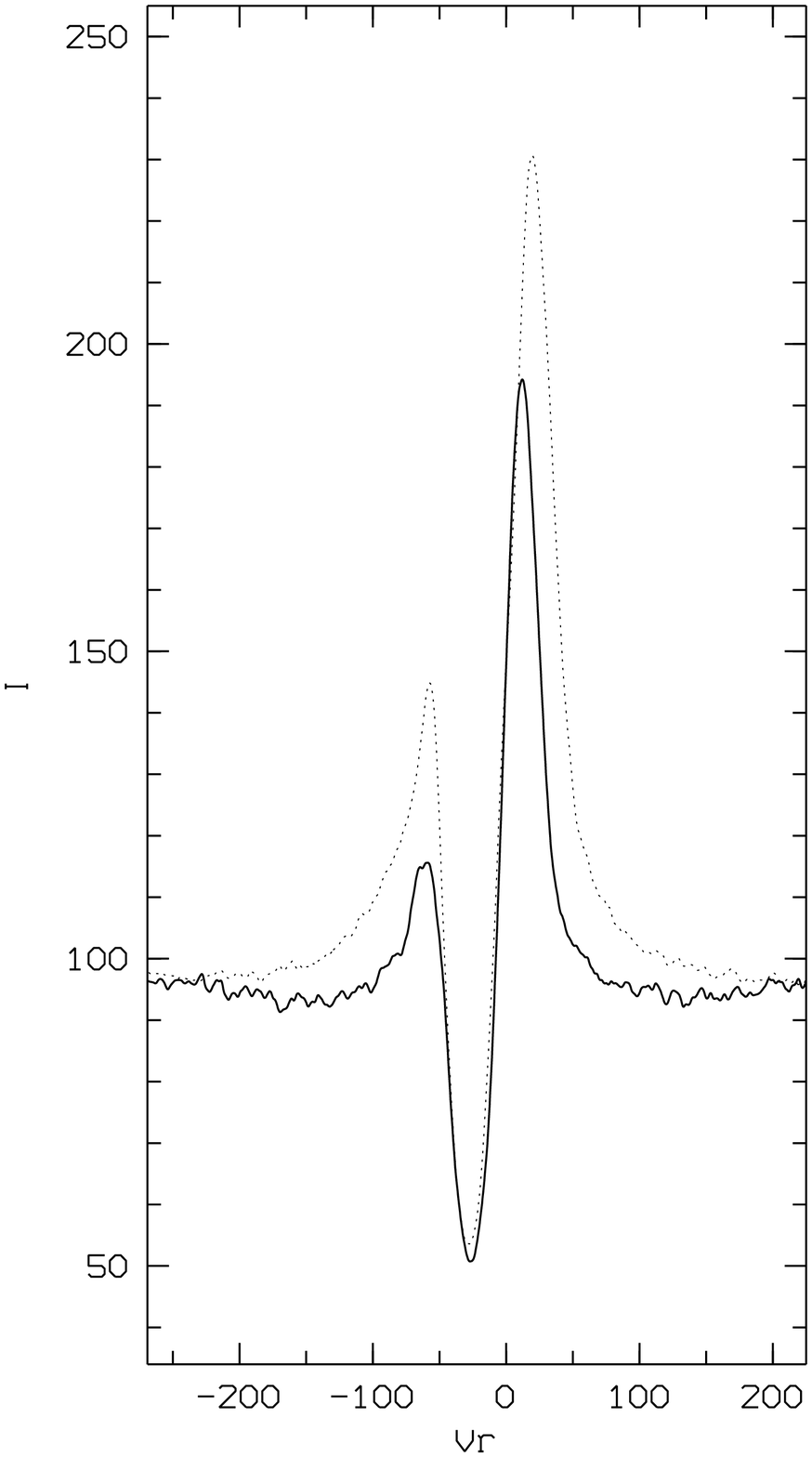}
\caption{H$\alpha$ line profile in the spectra of BD\,+48${\rm ^o}$1220
obtained on March 8, 2004 (solid curve) and January 10, 2004 (dashed curve).
The horizontal axis plots radial velocity, the continuum level is specified
to be 100. }
\end{figure}

\begin{figure}[t]	      		      
\includegraphics[angle=0,width=0.5\textwidth,bb=140 20 550 790,clip]{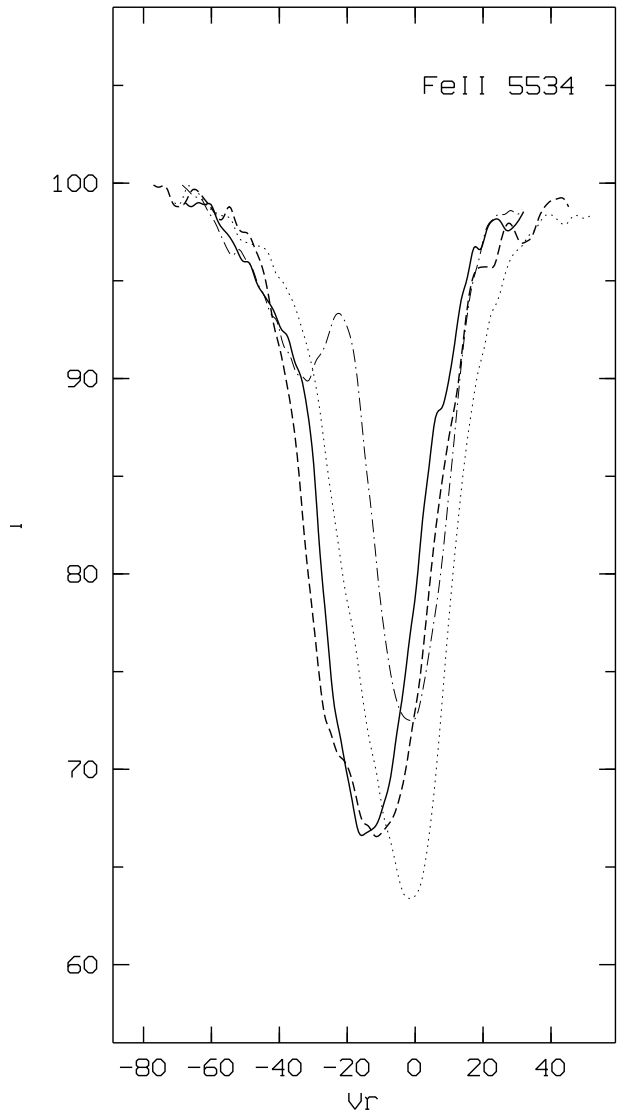}
\caption{FeII\,5534\,\AA{} line profile in the spectra of BD\,+48${\rm ^o}$1220
        obtained on December 2, 2002 (solid curve), September 9, 2003 (dotted curve),
        January 10, 2004 (dashed curve), and March 8, 2004 (dash--dotted curve).
	The horizontal axis plots radial velocity, the continuum level is specified
to be 100. }
\end{figure}

\begin{figure}[t]	      		      
\includegraphics[angle=0,width=0.7\textwidth,bb=140 20 550 790,clip]{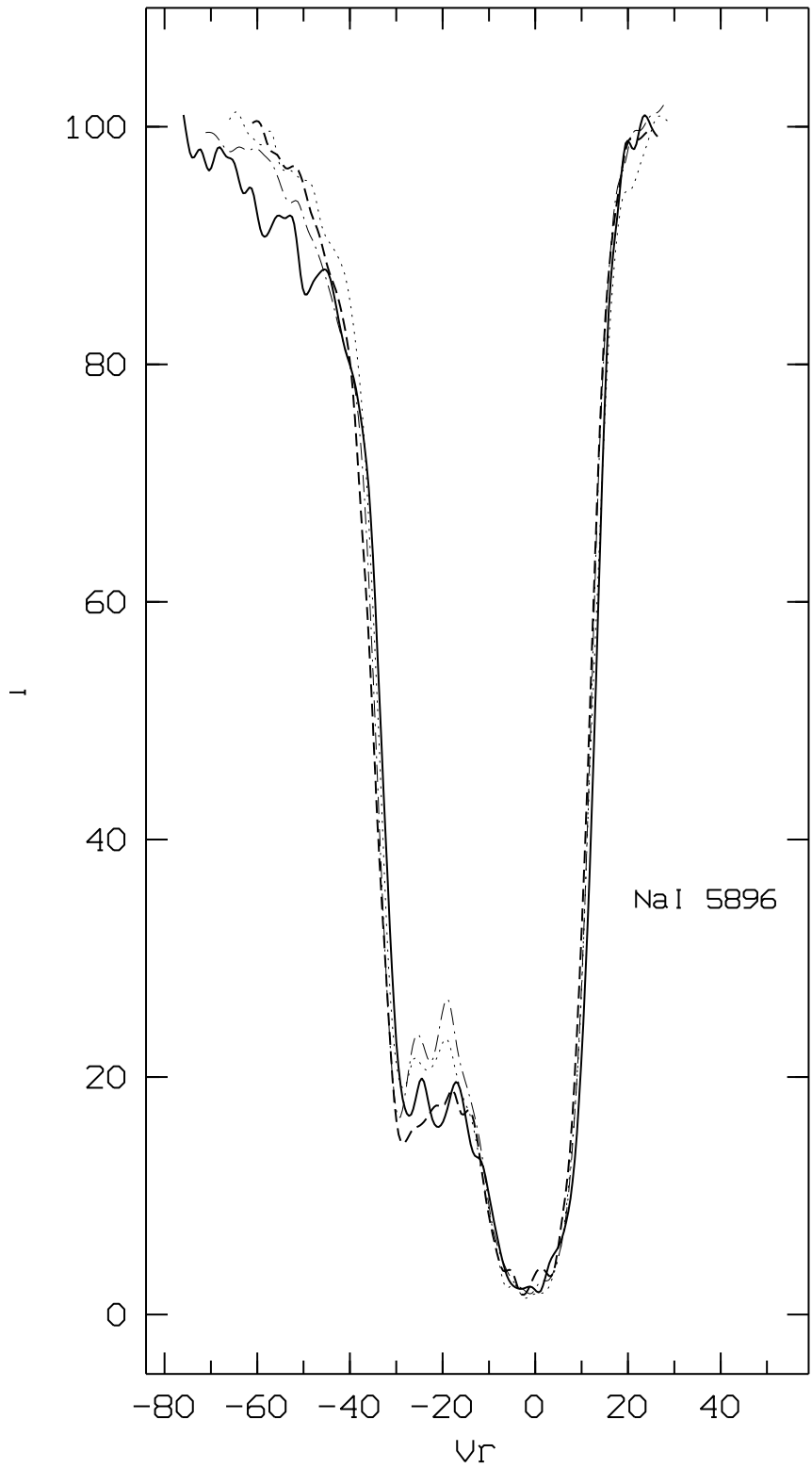}
\caption{Same as Fig.\,2 for the Na D1--line 5896\,\AA{}. }
\end{figure}

\begin{figure}[t]	      		      
\includegraphics[angle=-90,width=1.0\textwidth,bb=30 110 560 790,clip]{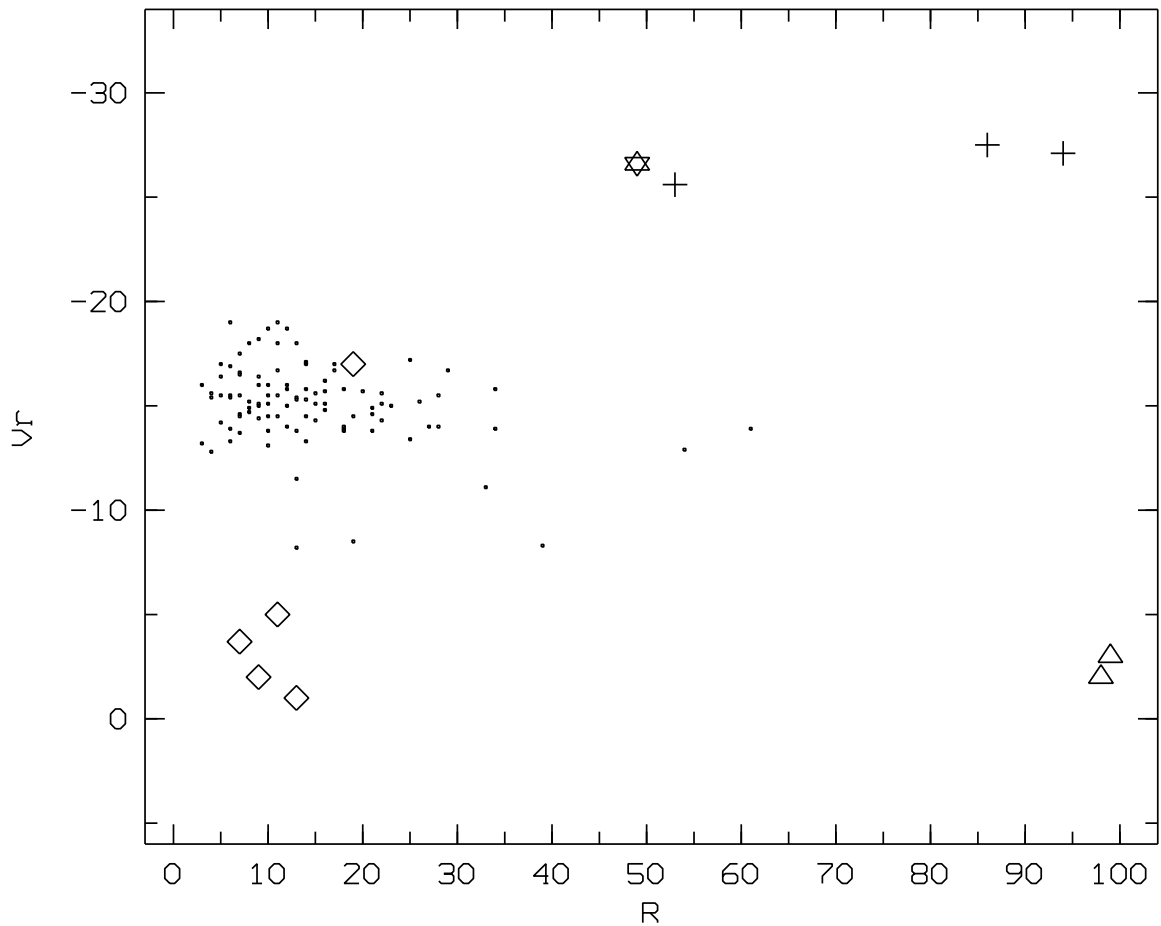}
\caption{Relation between the radial velocities and central depths derived
from lines in the spectrum of BD\,+48${\rm ^o}$1220 obtained on January
10, 2004. The asterisk marks the H$\alpha$ line, the crosses the short--wavelength
 D NaI components and the strong -- FeII(49)\,5316\,\AA{} line,
 the triangles the D NaI long--wavelength components,
 and the diamonds --  DIBs.}
\end{figure}

\begin{figure}[t]	      		      
\includegraphics[angle=-90,width=0.7\textwidth,bb=40 120 560 790,clip]{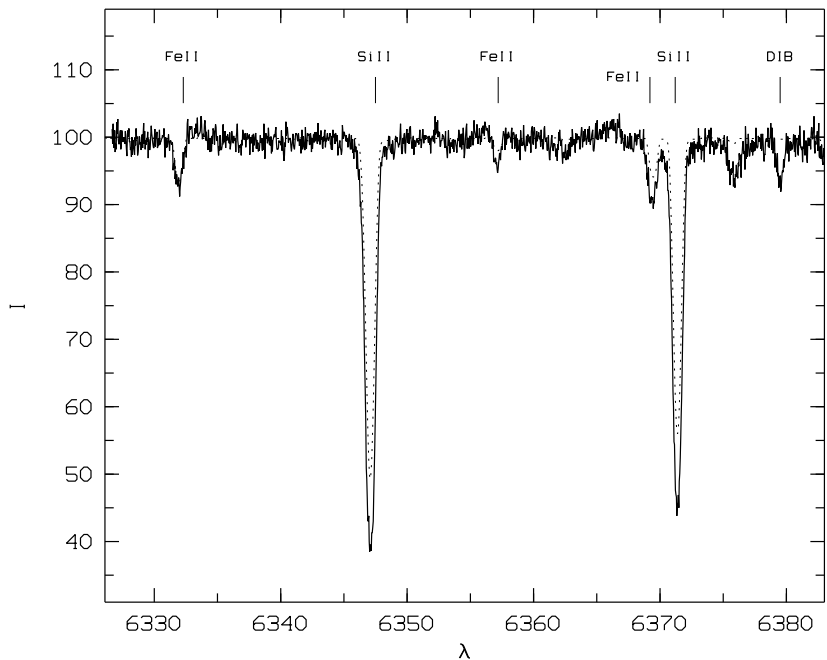}
\caption{Fragment of the observed spectrum of BD\,+48${\rm ^o}$1220 in the vicinity
of the SiII\,6347, 6371\,\AA{} lines (solid curve; the spectrum was obtained on
March 8, 2004) together with the theoretical spectrum calculated with the
model parameters $T_{eff}$=7900\,K, $\log g$=0.5, $\xi_t$=6.0\,km/s and
the chemical composition from Tabl.\,3 (dotted curve).}
\end{figure}

\begin{figure}[t]	      		      
\includegraphics[angle=-90,width=0.7\textwidth,bb=40 120 560 790,clip]{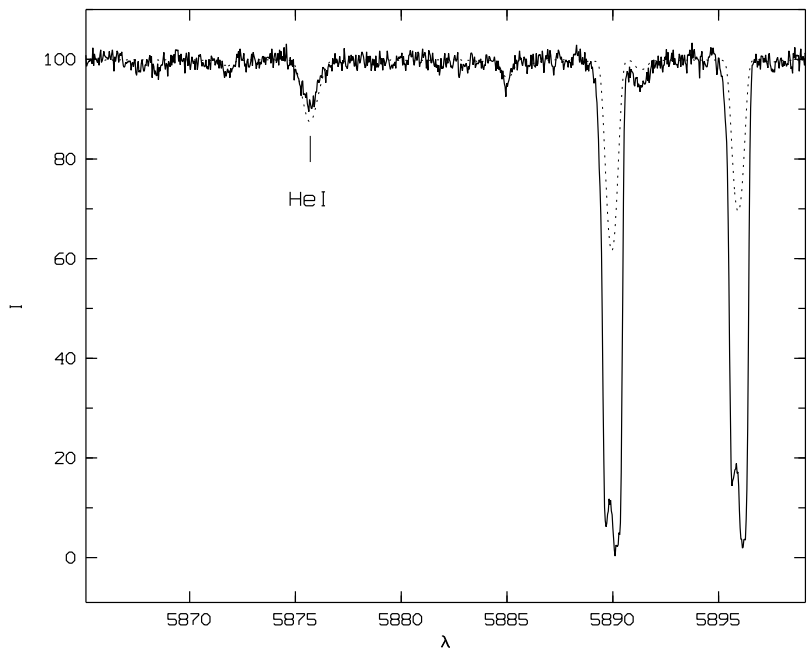}
\caption{Same as Fig.\,5 in the vicinity of the HeI\,5876\,\AA{} line.}
\end{figure}

\begin{figure}[t]	      		      
\includegraphics[angle=-90,width=0.7\textwidth,bb=40 120 560 790,clip]{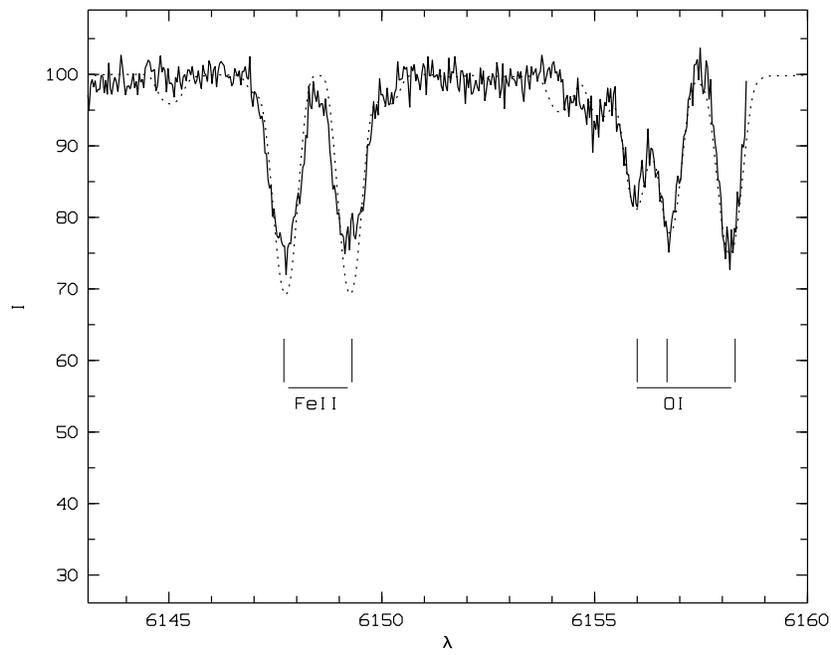}
\caption{Same as Fig.\,5 in the region of the OI\,6155\,\AA{} lines.}
\end{figure}

\clearpage
\newpage
\begin{table}[h!]
\caption{Observations of BD\,+48${\rm ^o}$1220 with the 6-m telescope}
\begin{tabular}{c@{\quad}|c @{\quad}|c@{\quad}|c|r}
\hline
Date  & Integration  & $\Delta \lambda$ & R   & S/N\\
     & time, s & \AA{}      & &    \\
\noalign{\smallskip}
\hline\noalign{\smallskip}
02.12.2002 & 2$\times$3600 & 4520--6000 & 60000 & 75  \\
09.09.2003 & 2$\times$3600 & 4520--6000 & 60000 & 100  \\
10.01.2004 & 2$\times$3600 & 5280--6760 & 75000 & 85  \\
08.03.2004 & 2$\times$3600 & 5280--6760 & 75000 & 115  \\
\hline
\end{tabular}
\end{table}

\begin{table}[hbtp]					    
\caption{Diffuse bands in the spectrum of BD\,+48${\rm ^o}$1220}
\medskip						    
\label{DIB}						    
\begin{tabular}{l@{\qquad}|r}			    
\hline							    
$\lambda$, \AA{} & $W_{\lambda}$, m\AA{}    \\
\hline 
5747 & 42  \\ 
5780 & 227 \\
5797 & 80  \\
6195 & 25  \\
6203 & 64  \\
6270 & 64  \\
6379 & 54  \\
6425 & 13  \\
6613 & 104 \\
6699 & 17  \\
\hline
\end{tabular}
\end{table}

\clearpage
\newpage
\begin{center}
\begin{table}[h]
\caption{Average heliocentric radial velocities for line groups and some
individual lines, from spectra of BD\,+48${\rm ^o}$1220 obtained on different dates.
Uncertain values are given by italic} 
\bigskip	 
\begin{tabular}{l @{\qquad}|  l @{\quad}| l @{\quad}| l @{\quad}| l}
\hline
Spectral  & \multicolumn{4}{c}{$V_{\odot}$ } \\
\cline{2-5}
details &02.12.02    &09.09.03    &10.01.04 &08.03.04     \\
\hline
\multicolumn{5}{l}{Absorption cores:} \\
FeII, CrII, É Ô.Ð. (${\rm r \rightarrow 1}$)& $-14.8$ & $-6.8$ & $-15.3$ & $-8.2$    \\
H$\alpha$             &            &            & $-27$   & $-28$        \\    
H$\beta$              &$-14.8$     & $-10.6$    &         &              \\
\hline
\multicolumn{5}{l}{Absorbtion line wings at r=0.90$ \div 0.95$}             \\
                      &$-14$       & $-9.5$     &$-15.5$  & $-13.0$      \\  
\hline 
\multicolumn{5}{l}{Maximum extension of short--wavelength wings:}\\
Fe\,{\sc II}(42)      & $-70$      & $-61$      &$-{\it 65}$& $-70$        \\
H$\alpha$             &            &            &$-{\it 65}$& $-{\it 57}$   \\    
\hline
\multicolumn{5}{l}{Emissions:} \\
Fe\,{\sc II}(40, 46)  &            &            &           & $-18$         \\
H$\alpha$             &            &            & $-15$       & $-{\it 13}$   \\
\hline
\multicolumn{5}{l}{Interstellar lines and bands:} \\
NaD       &$-25$, $-3$ &$-26$, $-2$ &$-27$, $-2.5$& $-28$, $-2$   \\
DIB                   &            &            &$-{\it 3}$   & $-{\it 2.5}$  \\     
\hline
\end{tabular}
\end{table}
\end{center}

\begin{table}[ht]
\caption{Elemental abundances $\log\varepsilon(X)$ (for $\log\varepsilon(H)$=12.0)
         in the atmosphere of BD\,+48${\rm ^o}$1220.  Solar abundances are taken
             from [\cite{Lodders}]}						    
\medskip						    
\begin{tabular}{l|c@{\quad}|l|r@{\quad}|r|r@{\quad}|c}			    
\hline							    
\multicolumn{2}{c|}{Sun}  &  \multicolumn{5}{c}{BD\,$+48^{\rm o}$\,1220, \quad 7900, 0.0, 6.0} \\
\cline{1-7}
E & $\log \varepsilon(E)$ &$X$ & $\log \varepsilon(X)$ &  $\sigma$ &\quad  n \quad &  $[X/Fe]_{\odot}$ \\
\hline
He    &10.90 & HeI &11.84 &      &1  & +1.04   \\ 
Li     &3.28 & LiI & 3.80 &      &1  & +0.62   \\ 
C      &8.39 & CI  & 8.38 &      &1  & +0.09    \\
O      &8.69 & OI  & 9.31 &  0.17&3  & +0.72   \\
Na     &6.30 & NaI & 7.07 &  0.31&3  & +0.87   \\
Mg     &7.55 & MgI & 7.14 &  0.20&3  &$-0.31$   \\
Si     &7.54 & SiII& 8.20 &      &1  & +0.76   \\
       &     &     & 8.25 &  0.16&3  & +0.81   \\
Ca     &6.34 & CaI & 6.04 &      &2  &$-0.20$  \\                                          
Sc     &3.07 & ScII& 2.63 &  0.28&4  &$-0.34$  \\
Ti     &4.92 & TiII& 4.42 &  0.19&23 &$-0.40$   \\
V      &4.00 & VII & 3.99 &  0.03&3  & +0.09   \\
Cr     &5.65 & CrI & 5.35 &  0.12&3  & $-0.20$  \\
       &     & CrII& 5.76 &  0.26&30 & +0.21   \\
Mn     &5.50 & MnI & 5.77 &      &1  & +0.37   \\
       &     & MnII& 5.57 &      &1  & +0.17   \\
Fe     &7.47 & FeI & 7.35 &  0.23&41 &$-0.07$  \\
       &     & FeII& 7.44 &  0.23&68 & +0.07   \\
Ni     &6.22 & NiI & 6.86 &  0.04&3  & +0.74   \\
Zn     &4.63 & ZnI & 4.97 &      &2  & +0.44   \\ 
Ba     &2.18 & BaII& 1.24 &      &2  &$-0.84$  \\
Eu     &0.52 & EuII&$-0.61$&     &1  &$-1.03$  \\
\hline
\end{tabular}
\label{chem}						    
\end{table}

\clearpage
\newpage

\section*{Appendix}

\bigskip
\bigskip

\setcounter{table}{0}

{\footnotesize

\begin{center}
\tablecaption{Residual line intensities $r$  in fractions of the continuum and 
       $V_{\odot}$  in the spectra of BD\,$+48^{\rm o}$\,1220 obtained on various dates.
       Uncertain values are given by italic}       
\tablehead{\hline Element&$\lambda$, \AA{} &\multicolumn{2}{c}{\underline{02.12.02}} &\multicolumn{2}{c}{\underline{09.09.03}} &\multicolumn{2}{c}{\underline{10.01.04}} &\multicolumn{2}{c}{\underline{08.03.04}}\\ 
      (multiplet) &   & $r$ & $V_{\odot}$  & $r$ & $V_{\odot}$ & $r$ & $V_{\odot}$ & $r$ & $V_{\odot}$ \\
    \hline \rule{0pt}{5pt}&&&&&&&&&\\ } 
\tabletail{\hline  }
\begin{supertabular}{l| l |r| r |r| r  |r| r  |r |r}
FeI(68)  &  4528.61 & 0.73 &$-$17.3      &  0.76 & $-$11.2 &&&&\\[5pt]
TiII(82) &  4529.49 & 0.72 & $-$13       &  0.71 & $-$6.5  &&&&\\[5pt]
TiII(50) &  4534.02:& 0.37 & --          &  0.35 &  --     &&&&\\
FeII(37) &  4534.16 &      &             &       &         &&&&\\ [5pt] 
CrII(39) &  4539.62 & 0.80 &$-$15.5      &  0.79 & $-$6.5  &&&&\\ [5pt]
FeII(38) &  4541.52 & 0.49 &$-$13.7      &  0.49 & $-$5.0  &&&&\\ [5pt]
TiII(60) &  4544.01 & 0.87 &$-$13.0      &  0.87 & $-$3.5  &&&&\\ [5pt]
TiII(30) &  4545.13 & 0.81 &$-$13.5      &  0.81 & $-$4.5  &&&&\\
\end{supertabular}
\end{center}
} 	   									    
\bigskip
\bigskip

{\footnotesize
\begin{center}
\tablecaption{Atomic parameters of the lines in the spectrum of
              BD\,+48${\rm ^o}$1220 obtained on January 10, 2004, and the corresponding
	      elemental abundances $\log \varepsilon(X)$ (for $\log \varepsilon(H)$=12)
	      in the atmosphere of  BD\,$+48^{\rm o}$\,1220 
	      obtained with parameters $T_{eff}=7900$\,K, $\log g$=0.5, $\xi_t=6.0$. }
\tablehead{\hline Element/$\lambda$&   EP, ev  & $\log gf$& $W_{\lambda}$& $\log \varepsilon(X)$  \\
    \hline 
    } 
\tabletail{\hline  }
\begin{supertabular}{c@{\quad}|r@{\quad}|@{\quad}r@{\quad}|r@{\quad}|r}
HeI       &          &        &      &            \\
5875.62   & 20.964   & 0.409  &   75 &  0.12      \\
LiI       &          &        &      &            \\
6707.76   & 0.000    &$-$0.009&   10 & $-$8.39      \\
CI        &          &        &      &            \\
5039.050  &   7.4828 &$$-$$2.100 &    23& $-$3.77      \\ 
OI        &          &        &      &            \\ 
  6155.966&  10.7400 &$$-$$1.011 &   117& $-$2.54      \\
\end{supertabular}
\end{center}  	   									   

\end{document}